\documentclass[letterpaper]{egu}

\usepackage{times}
\usepackage{graphicx}

\begin{document}


\title{Earthquake forecasting and its verification}
\runningtitle{Earthquake Forecasting}

\runningauthor{J.~R.~Holliday {\it et al.\/}}
\correspondence{J.~R.~Holliday (holliday@cse.ucdavis.edu)}

\author[1,2]{James R. Holliday}
\author[3,2]{Kazuyoshi Z. Nanjo}
\author[4  ]{Kristy F. Tiampo}
\author[2,1]{John B. Rundle}
\author[5  ]{Donald L. Turcotte}

\affil[1]{Department of Physics - University of California, Davis, USA}
\affil[2]{Computational Science and Engineering Center - University of
          California, Davis, USA}
\affil[3]{The Institute of Statistical Mathematics - Tokyo, Japan}
\affil[4]{Department of Earth Sciences - University of Western
          Ontario, Canada}
\affil[5]{Geology Department - University of California, Davis, USA}

\journal{\NPG}
\date{Manuscript version from 5 August 2005}

\firstauthor{Holliday}

\msnumber{npg-2005-0052}


\firstpage{1}

\maketitle


\begin{abstract}
No proven method is currently available for the reliable short time
prediction of earthquakes (minutes to months). However, it is possible
to make probabilistic hazard assessments for earthquake risk.  These
are primarily based on the association of small earthquakes with
future large earthquakes.  In this paper we discuss a new approach to
earthquake forecasting. This approach is based on a pattern
informatics (PI) method which quantifies temporal variations in
seismicity. The output is a map of areas in a seismogenic region
(``hotspots'') where earthquakes are forecast to occur in a future
10-year time span. This approach has been successfully applied to
California, to Japan, and on a worldwide basis.  These forecasts are
binary--an earthquake is forecast either to occur or to not occur.
The standard approach to the evaluation of a binary forecast is the
use of the relative operating characteristic (ROC) diagram, which is a
more restrictive test and less subject to bias than maximum likelihood
tests.  To test our PI method, we made two types of retrospective
forecasts for California.  The first is the PI method and the second
is a relative intensity (RI) forecast based on the hypothesis that
future earthquakes will occur where earthquakes have occurred in the
recent past.  While both retrospective forecasts are for the ten year
period 1 January 2000 to 31 December 2009, we performed an interim
analysis 5 years into the forecast.  The PI method out performs the RI
method under most circumstances.
\end{abstract}


\introduction

Earthquakes are the most feared of natural hazards because they occur
without warning. Hurricanes can be tracked, floods develop gradually,
and volcanic eruptions are preceded by a variety of precursory
phenomena. Earthquakes, however, generally occur without any warning.
There have been a wide variety of approaches applied to the
forecasting of earthquakes \citep{Mogi85, Turcotte91, Lomnitz94,
KeilisBorok02, Scholz02, Kanamori03}. These approaches can be divided
into two general classes; the first is based on empirical observations
of precursory changes. Examples include precursory seismic activity,
precursory ground motions, and many others. The second approach is
based on statistical patterns of seismicity. Neither approach has been
able to provide reliable short-term forecasts (days to months) on a
consistent basis.

Although short-term predictions are not available, long-term
seismic-hazard assessments can be made.  A large fraction of all
earthquakes occur in the vicinity of plate boundaries, although some
do occur in plate interiors.  It is also possible to assess the
long-term probability of having an earthquake of a specified magnitude
in a specified region. These assessments are primarily based on the
hypothesis that future earthquakes will occur in regions where past
earthquakes have occurred \citep{Frankel95, KossobokobKTM00}.
Specifically, the rate of occurrence of small earthquakes in a region
can be analyzed to assess the probability of occurrence of much larger
earthquakes.

The principal focus of this paper is a new approach to earthquake
forecasting \citep{RundleTKM02, TiampoRMK02, TiampoRMG02,
RundleTSKS03}.  Our method does not predict earthquakes, but it does
forecast the regions (hotspots) where earthquakes are most likely to
occur in the relatively near future (typically ten years). The
objective is to reduce the areas of earthquake risk relative to those
given by long-term hazard assessments. Our approach is based on
pattern informatics (PI), a technique that quantifies temporal
variations in seismicity patterns. The result is a map of areas in a
seismogenic region (hotspots) where earthquakes are likely to occur
during a specified period in the future. A forecast for California was
published by our group in 2002 \citep{RundleTKM02}.  Subsequently,
fifteen of the seventeen California earthquakes with magnitudes
M$\ge$5 occurred in or immediately adjacent to the resulting
hotspots. A forecast for Japan, presented in Tokyo in early October
2004, successfully forecast the location of the M$=$6.8 Niigata
earthquake that occurred on 23 October 2004.  A global forecast,
presented at the early December 2004 meeting of the American
Geophysical Union, successfully forecast the locations of the 23
December 2004, M$=$8.1 Macquarie Island earthquake, and the 26
December 2004 M$=$9.0 Sumatra earthquake. Before presenting further
details of these studies we will give a brief overview of the current
state of earthquake prediction and forecasting.


\section{Empirical approaches}

Empirical approaches to earthquake prediction rely on local
observations of precursory phenomena in the vicinity of the earthquake
to be predicted. It has been suggested that one or more of the
following phenomena may indicate a future earthquake \citep{Mogi85,
Turcotte91, Lomnitz94, KeilisBorok02, Scholz02, Kanamori03}: 1)
precursory increase or decrease in seismicity in the vicinity of the
origin of a future earthquake rupture, 2) precursory fault slip that
leads to surface tilt and/or displacements, 3) electromagnetic
signals, 4) chemical emissions, and 5) changes in animal behavior.

Examples of successful near-term predictions of future earthquakes
have been rare. A notable exception was the prediction of the M$=$7.3
Haicheng earthquake in northeast China that occurred on 4 February
1975. This prediction led to the evacuation of the city which
undoubtedly saved many lives. The Chinese reported that the successful
prediction was based on foreshocks, groundwater anomalies, and animal
behavior. Unfortunately, a similar prediction was not made prior to
the magnitude M$=$7.8 Tangshan earthquake that occurred on 28 July
1976 \citep{Utsu03}. Official reports placed the death toll in this
earthquake at 242,000, although unofficial reports placed it as high
as 655,000.

In order to thoroughly test for the occurrence of direct precursors
the United States Geological Survey (USGS) initiated the Parkfield
(California) Earthquake Prediction Experiment in 1985 \citep{BakunL85,
Kanamori03}. Earthquakes on this section of the San Andreas had
occurred in 1857, 1881, 1901, 1922, 1934, and 1966. It was expected
that the next earthquake in this sequence would occur by the early
1990's, and an extensive range of instrumentation was installed. The
next earthquake in the sequence finally occurred on 28 September
2004. No precursory phenomena were observed that were significantly
above the background noise level. Although the use of empirical
precursors cannot be ruled out, the future of those approaches does
not appear to be promising at this time.


\section{Statistical and statistical physics approaches}

A variety of studies have utilized variations in seismicity over
relatively large distances to forecast future earthquakes. The
distances are large relative to the rupture dimension of the
subsequent earthquake. These approaches are based on the concept that
the earth's crust is an activated thermodynamic system
\citep{RundleTSKS03}. Among the evidence for this behavior is the
continuous level of background seismicity in all seismographic
areas. About a million magnitude two earthquakes occur each year on
our planet. In southern California about a thousand magnitude two
earthquakes occur each year.  Except for the aftershocks of large
earthquakes, such as the 1992 M$=$7.3 Landers earthquake, this seismic
activity is essentially constant over time. If the level of background
seismicity varied systematically with the occurrence of large
earthquakes, earthquake forecasting would be relatively easy. This,
however, is not the case.

There is increasing evidence that there are systematic precursory
variations in some aspects of regional seismicity. For example, it has
been observed that there is a systematic variation in the number of
magnitude M$=$3 and larger earthquakes prior to at least some
magnitude M$=$5 and larger earthquakes, and a systematic variation in
the number of magnitude M$=$5 and larger earthquakes prior to some
magnitude M$=$7 and larger earthquakes.  The spatial regions
associated with this phenomena tend to be relatively large, suggesting
that an earthquake may resemble a phase change with an increase in the
``correlation length'' prior to an earthquake \citep{BowmanOSSS98,
JaumeS99}. There have also been reports of anomalous quiescence in the
source region prior to a large earthquake, a pattern that is often
called a ``Mogi Donut'' \citep{Mogi85, Kanamori03, WyssH88, Wyss97}.

Many authors have noted the occurrence of a relatively large number of
intermediate-sized earthquakes prior to a great earthquake. A specific
example was the sequence of earthquakes that preceded the 1906 San
Francisco earthquake \citep{SykesJ90}. This seismic activation has
been quantified as a power law increase in seismicity prior to
earthquakes \citep{BowmanOSSS98, JaumeS99, BufeV93, BufeNV94,
BrehmB98, BrehmB99, Main99, Robinson00, BowmanK01, YangVL01, KingB03,
BowmanS04, SammisBK04}. Unfortunately the success of these studies has
depended on knowing the location of the subsequent earthquake.

A series of statistical algorithms to make intermediate term
earthquake predictions have been developed by a Russian group under
the direction of V.~I.~Keilis-Borok using pattern recognition
techniques \citep{KeilisBorok90, KeilisBorok96}. Seismicity in
circular regions with diameters of 500~\unit{km} was analyzed. Based
primarily on seismic activation, earthquake alarms were issued for one
or more regions, with the alarms generally lasting for three years.
Alarms have been issued regularly since the mid 1980's and scored two
notable successes: the prediction of the 1988 Armenian earthquake and
the 1989 Loma Prieta earthquake. While a reasonably high success rate
has been achieved, there have been some notable misses including the
recent M$=$9.0 Sumatra and M$=$8.1 Macquerie Island earthquakes.

More recently, this group has used chains of premonitory earthquakes
as the basis for issuing alarms \citep{ShebalinKZUNT04,
KeilisBorokSGT04}. This method successfully predicted the M$=$6.5, 22
December 2003 San Simeon (California) earthquake and the M$=$8.1, 25
September 2003 Tokachi-oki, (Japan) earthquake with lead times of six
and seven months respectively. However, an alarm issued for southern
California, valid during the spring and summer of 2004, was a false
alarm.


\section{Chaos and forecasting}

Earthquakes are caused by displacements on preexisting faults.  Most
earthquakes occur at or near the boundaries between the near-rigid
plates of plate tectonics. Earthquakes in California are associated
with the relative motion between the Pacific plate and the North
American plate. Much of this motion is taken up by displacements on
the San Andreas fault, but deformation and earthquakes extend from the
Rocky Mountains on the east into the Pacific Ocean adjacent to
California on the west. Clearly this deformation and the associated
earthquakes are extremely complex.

It is now generally accepted that earthquakes are examples of
deterministic chaos \citep{Turcotte97}. Some authors
\citep{GellerJKM97, Geller97} have argued that this chaotic behavior
precludes the prediction of earthquakes. However, weather is also
chaotic, but forecasts can be made.  Weather forecasts are
probabilistic in the sense that weather cannot be predicted
exactly. One such example is the track of a hurricane. Probabilistic
forecasts of hurricane tracks are routinely made; sometimes they are
extremely accurate while at other times they are not. Another example
of weather forecasting is the forecast of El Ni\~no
events. Forecasting techniques based on pattern recognition and
principle components of the sea surface temperature fluctuation time
series have been developed that are quite successful in forecasting
future El Ni\~nos, but again they are probabilistic in nature
\citep{ChenCKZH04}. It has also been argued \citep{SykesSS99} that
chaotic behavior does not preclude the probabilistic forecasting of
future earthquakes. Over the past five years our group has developed
\citep{RundleTKM02, TiampoRMK02, TiampoRMG02, RundleTSKS03,
HollidayRTKD05} a technique for forecasting the locations where
earthquakes will occur based on pattern informatics (PI). This type of
approach has close links to principle component analysis, which has
been successfully used for the forecasting of El Ni\~nos.


\section{The PI method}

Seismic networks provide the times and locations of earthquakes over a
wide range of scales. One of the most sensitive networks has been
deployed over southern California and the resulting catalog is readily
available. Our objective has been to analyze the historical seismicity
for anomalous behavior that would provide information on the
occurrence of future earthquakes. At this point we are not able to
forecast the times of future earthquakes with precision. However, our
approach does appear to select the regions where earthquakes are most
likely to occur during a future time window. At the present time, this
time window is typically taken to be ten years, although it appears
that it is possible to utilize shorter time windows.

Our approach divides the seismogenic region to be studied into a grid
of square boxes whose size is related to the magnitude of the
earthquakes to be forecast. The rates of seismicity in each box are
studied to quantify anomalous behavior. The basic idea is that any
seismicity precursors represent changes, either a local increase or
decrease of seismic activity, so our method identifies the locations
in which these changes are most significant during a predefined change
interval. The subsequent forecast interval is the decadal time window
during which the forecast is valid.  The box size is selected to be
consistent with the correlation length associated with accelerated
seismic activity \citep{BowmanOSSS98}, and the minimum earthquake
magnitude considered is the lower limit of sensitivity and
completeness of the network in the region under consideration.

The detailed utilization of the PI method that we have used for
earthquake forecasting is as follows:

\begin{enumerate}

\item
The region of interest is divided into $N_B$ square boxes with linear
dimension $\Delta x$.  Boxes are identified by a subscript $i$ and are
centered at $x_i$.  For each box, there is a time series $N_i(t)$,
which is the number of earthquakes per unit time at time $t$ larger
than the lower cut-off magnitude $M_c$.  The time series in box $i$ is
defined between a base time $t_b$ and the present time $t$.

\item
All earthquakes in the region of interest with magnitudes greater than
a lower cutoff magnitude $M_c$ are included.  The lower cutoff
magnitude $M_c$ is specified in order to ensure completeness of the
data through time, from an initial time $t_0$ to a final time $t_2$.

\item
Three time intervals are considered:

\begin{enumerate}
  \item A reference time interval from $t_b$ to $t_1$.
  \item A second time interval from $t_b$ to $t_2$, $t_2 > t_1$.  The
        change interval over which seismic activity changes are
        determined is then $t_2 - t_1$.  The time $t_b$ is chosen to lie
        between $t_0$ and $t_1$.  Typically we take $t_0 = 1932$, $t_1
        = 1990$, and $t_2 = 2000$.  The objective is to quantify
        anomalous seismic activity in the change interval $t_1$ to
        $t_2$ relative to the reference interval $t_b$ to $t_1$.
  \item The forecast time interval $t_2$ to $t_3$, for which the
        forecast is valid.  We take the change and forecast intervals
        to have the same length.  For the above example, $t_3 = 2010$.
\end{enumerate}

\item
The seismic intensity in box $i$, $I_i(t_b,t)$, between two times
$t_b < t$, can then be defined as the average number of earthquakes with
magnitudes greater than $M_c$ that occur in the box per unit time
during the specified time interval $t_b$ to $t$.  Therefore, using
discrete notation, we can write:
\begin{equation}
  I_i(t_b,t) = \frac{1}{t-t_b} \sum_{t'=t_b}^t N_i(t'),
\end{equation}
where the sum is performed over increments of the time series, say days.

\item
In order to compare the intensities from different time intervals, we
require that they have the same statistical properties.  We therefore
normalize the seismic intensities by subtracting the mean seismic
activity of all boxes and dividing by the standard deviation of the
seismic activity in all boxes.  The statistically normalized seismic
intensity of box $i$ during the time interval $t_b$ to $t$ is then
defined by
\begin{equation}
  \hat{I}_i(t_b,t) = \frac{ I_i(t_b,t) - < I_i(t_b,t) >}{\sigma(t_b,t)},
\end{equation}
where $< I_i(t_b,t) >$ is the mean intensity averaged over all the boxes
and $\sigma(t_b,t)$ is the standard deviation of intensity over all
the boxes.

\item
Our measure of anomalous seismicity in box $i$ is the difference
between the two normalized seismic intensities:
\begin{equation}
  \Delta I_i(t_b,t_1,t_2) = \hat{I}_i(t_b,t_2) - \hat{I}_i(t_b,t_1).
\end{equation}

\item
To reduce the relative importance of random fluctuations (noise) in
seismic activity, we compute the average change in intensity,
$\overline{\Delta I_i(t_0,t_1,t_2)}$ over all possible pairs of
normalized intensity maps having the same change interval:
\begin{equation}
  \overline{\Delta I_i(t_0,t_1,t_2)} =
  \frac{1}{t_1 - t_0} \sum_{t_b=t_0}^{t_1} \Delta I_i(t_b,t_1,t_2),
\end{equation}
where the sum is performed over increments of the time series,
which here are days. 

\item
We define the probability of a future earthquake in box $i$,
$P_i(t_0,t_1,t_2,)$, as the square of the average intensity change:
\begin{equation}
  P_i(t_0,t_1,t_2,) = \overline{\Delta I_i(t_b,t_1,t_2)}^2.
\end{equation}

\item
To identify anomalous regions, we wish to compute the change in the
probability $P_i(t_0,t_1,t_2,)$ relative to the background so that we
subtract the mean probability over all boxes.  We denote this change
in the probability by
\begin{equation}
  \Delta P_i(t_0,t_1,t_2) = P_i(t_0,t_1,t_2) - < P_i(t_0,t_1,t_2) >,
\end{equation}
where $< P_i(t_0,t_1,t_2) >$ is the background probability for a large
earthquake.

\end{enumerate}

Hotspots are defined to be the regions where $\Delta P_i(t_0,t_1,t_2)$
is positive. In these regions, $P_i(t_0,t_1,t_2)$ is larger than the
average value for all boxes (the background level). Note that since
the intensities are squared in defining probabilities the hotspots may
be due to either increases of seismic activity during the change time
interval (activation) or due to decreases (quiescence). We hypothesize
that earthquakes with magnitudes larger than $M_c+2$ will occur
preferentially in hotspots during the forecast time interval $t_2$ to
$t_3$.


\section{Applications of the PI method}

\begin{figure}
  \centering
  \includegraphics[width=\columnwidth]{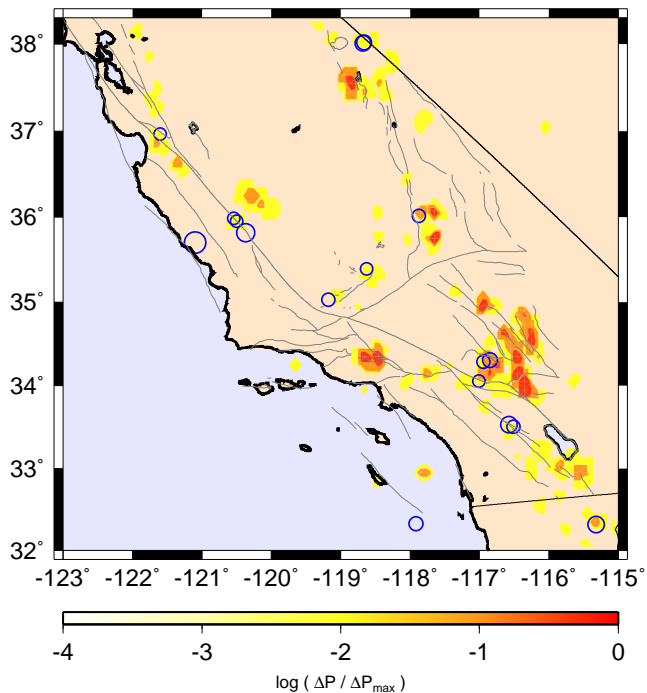}
  \caption{\label{fig.california}
  Application of the PI method to southern California. Colored areas
  are the forecast hotspots for the occurrence of M$\ge$5 earthquakes
  during the period 2000-2010 derived using the PI method. The color
  scale gives values of the $\log_{10}(P / P_{max})$. Also shown are
  the locations of the seventeen earthquakes with M$\ge$5 that have
  occurred in the region since 1 January 2000. Fifteen of the
  seventeen earthquakes were successfully forecast. More details of
  the earthquakes are given in Table~\ref{table.quakes}.}
\end{figure}

The PI method was first applied to seismicity in southern California
and adjacent regions ($32^\circ$ to $37^\circ$ N lat, $238^\circ$ to
$245^\circ$ E long). This region was divided into a grid of 3500 boxes
with $\Delta x=0.1^\circ$ (11~\unit{km}). Consistent with the
sensitivity of the southern California seismic network, the lower
magnitude cutoff was taken to be M$=$3. The initial time was
$t_0$=1932, the change interval was from $t_1$=1990 to $t_2$=2000, and
the forecast interval was from $t_2$=2000 to $t_3$=2010. The initial
studies for California were published in 2002 \citep{RundleTKM02}, the
results are reproduced in Figure~\ref{fig.california}. The colored
regions are the hotspots defined to be the boxes where $\Delta P$ is
positive. This forecast of where earthquakes would likely occur was
considered to be valid for the forecast interval from 2000 to 2010 and
would be applicable for earthquakes with M$=$5 and larger. Since 1
January 2000 seventeen earthquakes with M$\ge$5 have occurred in the
test region. These are also shown in Figure~\ref{fig.california}, and
information on these earthquakes is given in Table~\ref{table.quakes}.
We consider the forecast to be successful if the epicenter of the
earthquake lies within a hotspot box or in one of the eight adjoining
boxes \citep{Moore62}.  Fifteen of the seventeen earthquakes were
successfully forecast.

\begin{table}
\caption{\label{table.quakes}
Earthquakes with M$\ge$5 that occurred in the California test region
since 1 January 2000. Fifteen of these seventeen earthquakes were
successfully forecast.  The two missed events are marked with an
asterisk.}
\begin{center}
\begin{tabular}{llll}
   & {\bf Event}         & {\bf Magnitude} & {\bf Date}   \\
 1 & Big Bear I          & M$=$5.1         & 10 Feb.  2001 \\
 2 & Coso                & M$=$5.1         & 17 July  2001 \\
 3 & Anza I              & M$=$5.1         & 32 Oct.  2001 \\
 4 & Baja                & M$=$5.7         & 22 Feb.  2002 \\
 5 & Gilroy              & M$=$5.0         & 13 May.  2002 \\
 6 & Big Bear II         & M$=$5.4         & 22 Feb.  2003 \\
 7 & San Simeon$^\star$  & M$=$6.5         & 22 Dec.  2003 \\
 8 & San Clemente Island$^\star$  & M$=$5.2         & 15 June  2004 \\
 9 & Bodie I             & M$=$5.5         & 18 Sept. 2004 \\
10 & Bodie II            & M$=$5.4         & 18 Sept. 2004 \\
11 & Parkfield I         & M$=$6.0         & 18 Sept. 2004 \\
12 & Parkfield II        & M$=$5.2         & 18 Sept. 2004 \\
13 & Arvin               & M$=$5.0         & 29 Sept. 2004 \\
14 & Parkfield III       & M$=$5.0         & 30 Sept. 2004 \\
15 & Wheeler Ridge       & M$=$5.2         & 16 April 2005 \\
16 & Anza II             & M$=$5.2         & 12 June  2005 \\
17 & Yucaipa             & M$=$5.0         & 16 June  2005
\end{tabular}
\end{center}
\end{table}

\begin{figure}
  \centering
  \includegraphics[width=\columnwidth]{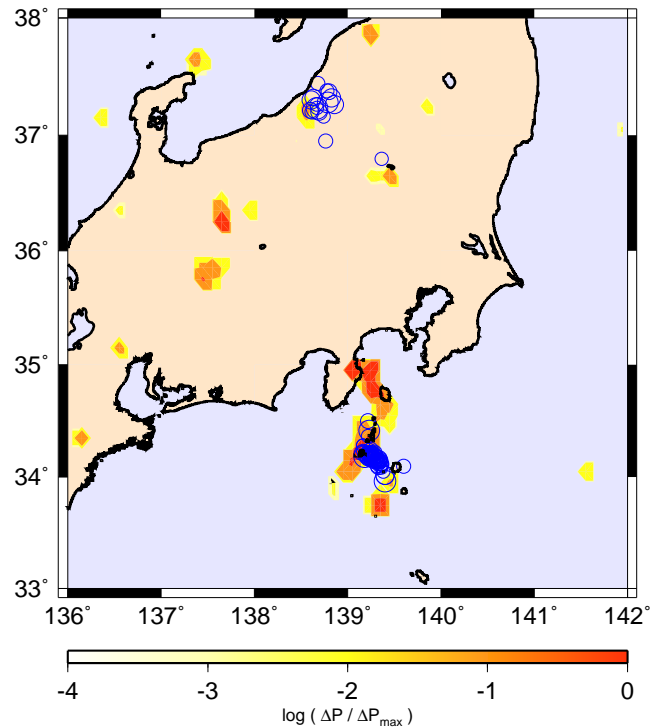}
  \caption{\label{fig.japan}
  Application of the PI method to central Japan. Colored
  areas are the forecast hotspots for the occurrence of M$\ge$5
  earthquakes during the period 2000-2010 derived using the PI
  method. The color scale gives values of the $\log_{10}(P /
  P_{max})$. Also shown are the locations of the 99 earthquakes with
  M$\ge$5 that have occurred in the region since 1 January 2000.}
\end{figure}

The second area to which the PI method was applied was Japan. The
forecast hotspots for the Tokyo region ($33^\circ$ to $38^\circ$ N
lat, $136^\circ$ to $142^\circ$ W long) are given in
Figure~\ref{fig.japan}. The initial time was $t_0$=1965, and the
change and forecast intervals were the same as those used for
California. Between 1 January 2000 and 14 October 2004, 99 earthquakes
occurred and 91 earthquakes were successfully forecast. This forecast
was presented at the International Conference on Geodynamics, 14-16
October 2004, Tokyo by one of the authors (JBR). Subsequently the
Niigata earthquake (M$=$6.8) occurred on 23 October 2004.  This
earthquake and its subsequent M$\ge$5 aftershocks were successfully
forecast.

\begin{figure}
  \centering
  \includegraphics[width=\columnwidth]{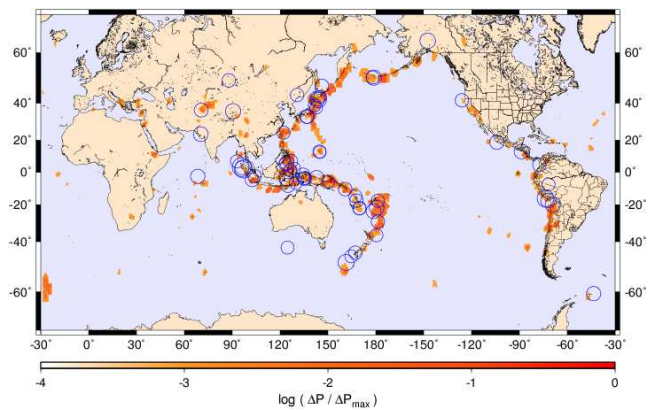}
  \caption{\label{fig.world}
  World-wide application of the PI method. Colored areas are
  the forecast hotspots for the occurrence of M$\ge$7 earthquakes
  during the period 2000-2010 derived using the PI method. The color
  scale gives values of the $\log_{10}(P / P_{max})$. Also shown are
  the locations of the sixty three earthquakes with M$\ge$7 that have
  occurred in the region since 1 January 2000.}
\end{figure}

The PI method has also been applied on a worldwide basis. In this case
$1^\circ\times1^\circ$ boxes were considered, $\Delta x=1^\circ$
(110~\unit{km}). Consistent with the sensitivity of the global seismic
network the lower magnitude cutoff was taken to be $M_c=5$. The
initial time was $t_0$=1965; the change and forecast intervals were
the same as above. The resulting map of hotspots was presented by two
of the authors (DLT and JRH) at the Fall Meeting of the American
Geophysical Union on 14 December 2004 (abstracts: AGUF2004NG24B-01 and
AGUF2004NG54A-08).  This map is given in Figure~\ref{fig.world}. This
forecast of where earthquakes would occur was considered to be valid
for the period 2000 to 2010 and would be applicable for earthquakes
with magnitudes greater than 7.0. Between 1 January 2000 and 14
December 2004 there were 63 M$\ge$7 earthquakes worldwide; 55 of these
earthquakes occurred within a hotspot or adjoining boxes. Subsequent
to the meeting presentation, the M$=$8.1 Macquarie Island earthquake
occurred on 23 December 2004 and the M$=$9.0 Sumatra earthquake
occurred on 26 December 2004. The epicenters of both earthquakes were
successfully forecast.

\section{Forecast verification}

Previous tests of earthquake forecasts have emphasized the likelihood
test \citep{KaganJ00, RundleTKM02, TiampoRMK02, HollidayRTKD05}.
These tests have the significant disadvantage that they are overly
sensitive to the least probable events.  For example, consider two
forecasts.  The first perfectly forecasts 99 out of 100 events but
assigns zero probability to the last event.  The second assigns zero
probability to all 100 events.  Under a likelihood test, both
forecasts will have the same skill score of $-\infty$.  Furthermore, a
naive forecast that assigns uniform probability to all possible sites
will always score higher than a forecast that misses only a single
event but is otherwise superior.  For this reason, likelihood tests
are more subject to unconscious bias.

An extensive review on forecast verification in the atmospheric
sciences has been given by \citet{JoliffeS03}.  The wide variety of
approaches that they consider are directly applicable to earthquake
forecasts as well.  The earthquake forecasts considered in this paper
can be viewed as binary forecasts by considering the events
(earthquakes) as being forecast either to occur or not to occur in a
given box.  We consider that there are four possible outcomes for each
box, thus two ways to classify each red, hotspot, box, and two ways to
classify each white, non-hotspot, box:

\begin{enumerate}
\item An event occurs in a hotspot box or within the Moore neighborhood
      of the box (the Moore neighborhood is comprised of the eight
      boxes surrounding the forecast box).  This is a success.
\item No event occurs in a white non-hotspot box.  This is also
      a success.
\item No event occurs in a hotspot box or within the Moore neighborhood of
      the hotspot box.  This is a false alarm.
\item An event occurs in a white, non-hotspot box.  This is a failure
      to forecast.
\end{enumerate}

We note that these rules tend to give credit, as successful forecasts,
for events that occur very near hotspot boxes.  We have adopted these
rules in part because the grid of boxes is positioned arbitrarily on
the seismically active region, thus we allow a margin of error of $\pm
1$ box dimension.  In addition, the events we are forecasting are
large enough so that their source dimension approaches, and can even
exceed, the box dimension meaning that an event might have its
epicenter outside a hotspot box, but the rupture might then propagate
into the box.  Other similar rules are possible but we have found that
all such rules basically lead to similar results.

The standard approach to the evaluation of a binary forecast is the
use of a relative operating characteristic (ROC) diagram
\citep{Swets73, Mason03}.  Standard ROC diagrams consider the fraction
of failures-to-predict and the fraction of false alarms.  This method
evaluates the performance of the forecast method relative to random
chance by constructing a plot of the fraction of failures to predict
against the fraction of false alarms for an ensemble of forecasts.
\citet{Molchan97} has used a modification of this method to evaluate
the success of intermediate term earthquake forecasts.

The binary approach has a long history, over 100 years, in the
verification of tornado forecasts \citep{Mason03}.  These forecasts
take the form of a tornado forecast for a specific location and time
interval, each forecast having a binary set of possible outcomes.  For
example, during a given time window of several hours duration, a
forecast is issued in which a list of counties is given with a
statement that one or more tornadoes will or will not occur.  A
$2\times2$ {\it contingency table\/} is then constructed, the top row
contains the counties in which tornadoes are forecast to occur and the
bottom row contains counties in which tornadoes are forecast to not
occur.  Similarly, the left column represents counties in which
tornadoes were actually observed, and the right column represents
counties in which no tornadoes were observed.

With respect to earthquakes, our forecasts take exactly this form.  A
time window is proposed during which the forecast of large earthquakes
having a magnitude above some minimum threshold is considered valid.
An example might be a forecast of earthquakes larger than $M=5$ during
a period of five or ten years duration.  A map of the seismically
active region is then completely covered (``tiled'') with boxes of two
types: boxes in which the epicenters of at least one large earthquake
are forecast to occur and boxes in which large earthquakes are
forecast to not occur.  In other types of forecasts, large earthquakes
are given some continuous probability of occurrence from 0\% to 100\%
in each box \citep{KaganJ00}.  These forecasts can be converted to the
binary type by the application of a {\it threshold\/} value.  Boxes
having a probability below the threshold are assigned a forecast
rating of {\it non-occurrence\/} during the time window, while boxes
having a probability above the threshold are assigned a forecast
rating of {\it occurrence\/}.  A high threshold value may lead to many
{\it failures to predict\/} (events that occur where no event is
forecast), but few {\it false alarms\/} (an event is forecast at a
location but no event occurs).  The level at which the threshold is
set is then a matter of public policy specified by emergency planners,
representing a balance between the prevalence of failures to predict
and false alarms.


\section{Binary earthquake forecast verification}

\begin{figure}
  {\bf (A)} \\
  \includegraphics[angle=270,width=\columnwidth]{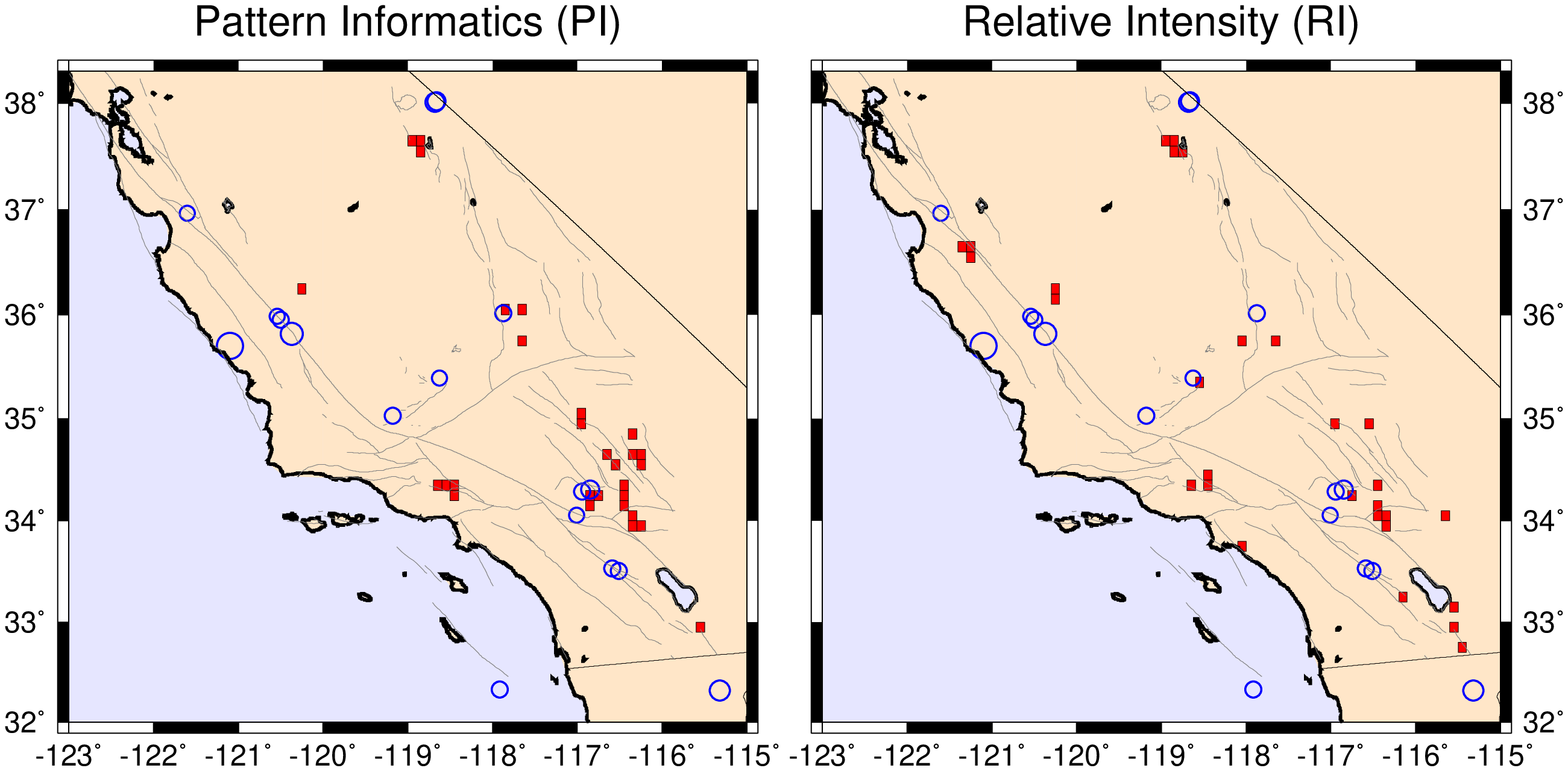}
  \vspace{2Em} \\
  {\bf (B)} \\
  \includegraphics[angle=270,width=\columnwidth]{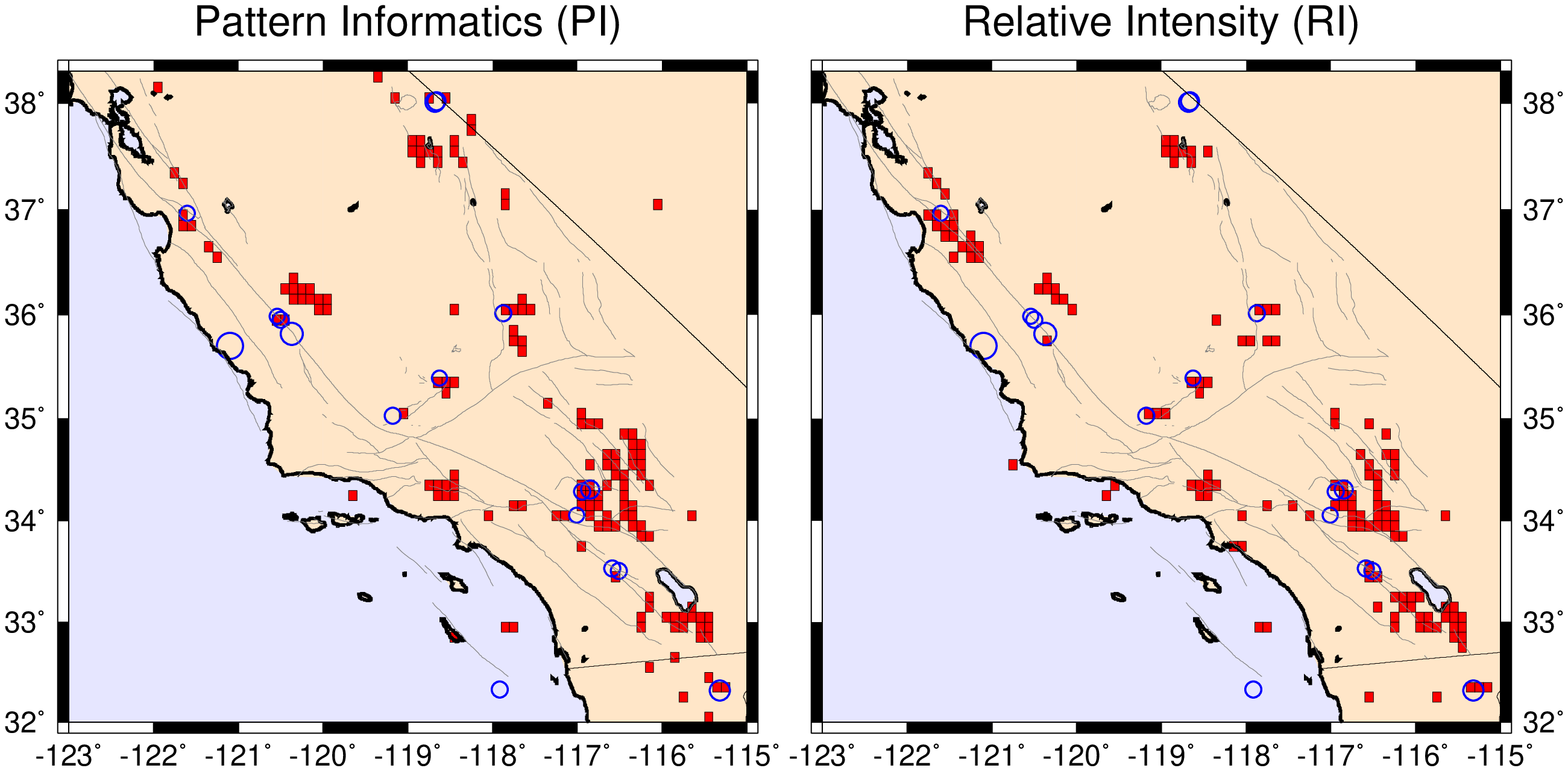}
  \caption{\label{fig.pixels}
  Retrospective application of the PI and RI methods for southern
  California as a function of false alarm rate.  Red boxes are the
  forecast hotspots for the occurrence of $M\ge5$ earthquakes during
  the period 2000 to 2005.  Also shown are the locations of the
  $M\ge5$ earthquakes that occurred in this region during the forecast
  period.  In Figure~\ref{fig.pixels}A, a threshold value was chosen
  such that $F \approx 0.005$.  In Figure~\ref{fig.pixels}B, a threshold value
  was chosen such that $F \approx 0.021$.}
\end{figure}

To illustrate this approach to earthquake forecast verification, we
have constructed two types of retrospective binary forecasts for
California.  The first type of forecast utilizes the PI method
described above.  We apply the method to southern California and
adjacent regions ($32^\circ$ to $38.3^\circ$ N lat, $238^\circ$ to
$245^\circ$ E long) using a grid of boxes with $\Delta x = 0.1^\circ$
and a lower magnitude cutoff $M_c = 3.0$.  For this retrospective
forecast we take the initial time $t_0=1932$, the change interval
$t_1=1989$ to $t_2=2000$, and the forecast interval $t_2=2000$ to
$t_3=2010$ \citep{RundleTKM02, TiampoRMK02}.  In the analysis given
above we considered regions with $\Delta P$ positive to be hotspots.
The PI forecast under the above conditions with $\Delta P > 0$ is
given in Figure~\ref{fig.pixels}B.  Hotspots include 127 of the 5040
boxes considered.  This forecast corresponds to that given in
Figure~\ref{fig.california}.  The threshold for hotspot activation can
be varied by changing the threshold value for $\Delta P$.  A forecast
using a higher threshold value is given in Figure~\ref{fig.pixels}A.
Hotspots here include only 29 of the 5040 boxes considered.

An alternative approach to earthquake forecasting is to use the rate
of occurrence of earthquakes in the past.  We refer to this type of
forecast as a {\it relative intensity\/} (RI) forecast.  In such a
forecast, the study region is tiled with boxes of size $0.1^\circ
\times 0.1^\circ$.  The number of earthquakes with magnitude $M\ge3.0$
in each box down to a depth of 20~\unit{km} is determined over the
time period from $t_0=1932$ to $t_2=2000$.  The RI score for each box
is then computed as the total number of earthquakes in the box in the
time period divided by the value for the box having the largest value.
A threshold value in the interval $[0,1]$ is then selected.  Large
earthquakes having $M\ge5$ are then considered possible only in boxes
having an RI value larger than the threshold.  The remaining boxes
with RI scores smaller than the threshold represent sites at which
large earthquakes are forecast to not occur.  The physical
justification for this type of forecast is that large earthquakes are
considered most likely to occur at sites of high seismic activity.

In order to make a direct comparison of the RI forecast with the PI
forecast, we select the threshold for the RI forecast to give the same
box coverage given for the PI forecast in Figures \ref{fig.pixels}A
and \ref{fig.pixels}B, {\it i.e.\/} 29 boxes and 127 boxes
respectively.  Included in all figures are the earthquakes with
$M\ge5$ that occurred between 2000 and 2005 in the region under
consideration.

\section{Contingency tables and ROC diagrams}

The first step in our generation of ROC diagrams is the construction
of the $2\times2$ contingency table for the PI and RI forecast maps
given in Figure~\ref{fig.pixels}.  The hotspot boxes in each map
represent the forecast locations.  A hotspot box upon which {\it at
least\/} one large future earthquake during the forecast period occurs
is counted as a {\it successful forecast\/}.  A hotspot box upon which
{\it no\/} large future earthquake occurs during the forecast period
is counted as an {\it unsuccessful forecast\/}, or alternately, a {\it
false alarm\/}.  A white box upon which {\it at least\/} one large
future earthquake during the forecast period occurs is counted as a
{\it failure to forecast\/}.  A white box upon which {\it no\/} large
future earthquake occurs during the forecast period is counted as a
{\it unsuccessful forecast of non-occurrence\/}.

Verification of the PI and RI forecasts proceeds in exactly the same
was as for tornado forecasts.  For a given number of hotspot boxes,
which is controlled by the value of the probability threshold in each
map, the contingency table (see Table~\ref{table.contingency}) is
constructed for both the PI and RI maps.  Values for the table
elements $a$ (Forecast=yes, Observed=yes), $b$ (Forecast=yes,
Observed=no), $c$ (Forecast=no, Observed=yes), and $d$ (Forecast=no,
Observed=no) are obtained for each map.  The fraction of colored
boxes, also called the {\it probability of forecast of occurrence\/},
is $r=(a+b)/N$, where the total number of boxes is $N=a+b+c+d$.  The
{\it hit rate\/} is $H=a/(a+c)$ and is the fraction of large
earthquakes that occur on a hotspot.  The {\it false alarm rate\/} is
$F=b/(b+d)$ and is the fraction of non-observed earthquakes that are
incorrectly forecast.

\begin{table}
\caption{\label{table.contingency}
Contingency tables as a function of false alarm rate.  In
Table~\ref{table.contingency}A, a threshold value was chosen such that
$F \approx 0.005$.  In Table~\ref{table.contingency}B, a threshold
value was chosen such that $F \approx 0.021$.}
{\bf (A)}
\begin{center}
\begin{tabular}{|c|c|c|c|}
\multicolumn{4}{c}{Pattern informatics (PI) forecast} \\ \hline
Forecast & \multicolumn{3}{c|}{Observed} \\ \cline{2-4}
         &  Yes    & No       & Total    \\ \hline
Yes      & (a) 4   & (b) 25   & 29       \\
No       & (c) 13  & (d) 4998 & 5011     \\ \hline
Total    &  17     & 5023     & 5040     \\ \hline
\end{tabular}
\begin{tabular}{|c|c|c|c|}
\multicolumn{4}{c}{Relative intensity (RI) forecast} \\ \hline
Forecast & \multicolumn{3}{c|}{Observed} \\ \cline{2-4}
         &  Yes    & No       & Total    \\ \hline
Yes      & (a) 2   & (b) 27   & 29       \\
No       & (c) 14  & (d) 4997 & 5011     \\ \hline
Total    &  16     & 5024     & 5040     \\ \hline
\end{tabular}
\end{center}
\vspace{2Em}

{\bf (B)}
\begin{center}
\begin{tabular}{|c|c|c|c|}
\multicolumn{4}{c}{Pattern informatics (PI) forecast} \\ \hline
Forecast & \multicolumn{3}{c|}{Observed} \\ \cline{2-4}
         &  Yes    & No       & Total    \\ \hline
Yes      & (a) 23  & (b) 104  & 127      \\
No       & (c) 9   & (d) 4904 & 4913     \\ \hline
Total    &  32     & 5008     & 5040     \\ \hline
\end{tabular}
\begin{tabular}{|c|c|c|c|}
\multicolumn{4}{c}{Relative intensity (RI) forecast} \\ \hline
Forecast & \multicolumn{3}{c|}{Observed} \\ \cline{2-4}
         &  Yes    & No       & Total    \\ \hline
Yes      & (a) 20  & (b) 107  & 127      \\
No       & (c) 10  & (d) 4903 & 4913     \\ \hline
Total    &  30     & 5010     & 5040     \\ \hline
\end{tabular}
\end{center}
\end{table}

\begin{figure}
  \centering
  \includegraphics[angle=270,width=\columnwidth]{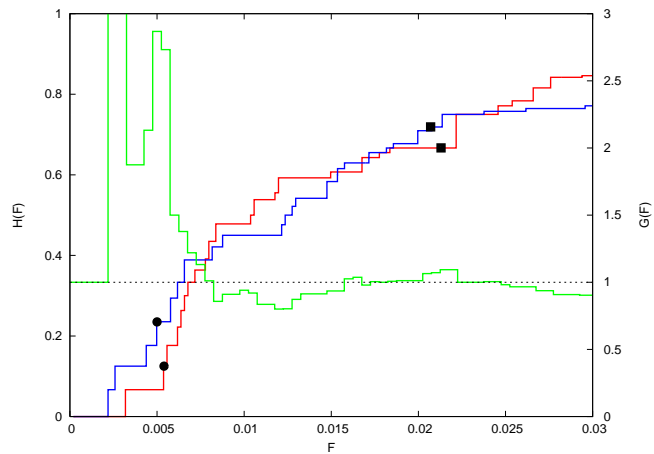}
  \caption{\label{fig.roc}
  Relative operating characteristic (ROC) diagram.  Plot of hit rates,
  $H$, versus false alarm rates, $F$, for the PI forecast (blue) and
  RI forecast (red).  Also shown is the gain ratio (green) defined as
  $H_{PI}(F) / H_{RI}(F)$.  The filled symbols correspond to the
  threshold values used in Figure~\ref{fig.pixels} and
  Table~\ref{table.contingency}, solid circles for 29 hotspot boxes
  and solid squares for 127 hotspot boxes.  The horizontal dashed line
  corresponds to zero gain.}
\end{figure}

To analyze the information in the PI and RI maps, the standard
procedure is to consider all possible forecasts together.  These are
obtained by increasing $F$ from 0 (corresponding to no hotspots on the
map) to 1 (all active boxes on the map are identified as hotspots).
The plot of $H$ versus $F$ is the relative operating characteristic
(ROC) diagram.  Varying the threshold value for both the PI and RI
forecasts, we have obtained the values of $H$ and $F$ given in
Figure~\ref{fig.roc}, blue for the PI forecasts and red for the RI
forecasts.  The results corresponding to the maps given in
Figure~\ref{fig.pixels} and the contingency tables given in
Table~\ref{table.contingency} are given by the filled symbols.
The forecast with 29 hotspot boxes (Figure~\ref{fig.roc}A and
Table~\ref{table.contingency}A) has $F_{PI} = 0.00498$, $H_{PI} =
0.235$ and $F_{RI} = 0.00537$, $H_{RI} = 0.125$.  The forecast with 127
hotspot boxes (Figure~\ref{fig.roc}B and
Table~\ref{table.contingency}B) has $F_{PI} = 0.0207$, $H_{PI} =
0.719$ and $F_{RI} = 0.0213$, $H_{RI} = 0.666$.  Also shown in
Figure~\ref{fig.roc} is a gain curve (green) defined by the ratio of
$H_{PI}(F)$ to $H_{RI}(F)$.  Gain values greater than unity indicate
better performance using the PI map than using the RI map.  The
horizontal dashed line corresponds to zero gain.  From
Figure~\ref{fig.roc} it can be seen that the PI approach outperforms
(is above) the RI under many circumstances and both outperform a
random map, where $H=F$, by a large margin.

\conclusions[Discussion]

The fundamental question is whether forecasts of the time and location
of future earthquakes can be accurately made. It is accepted that long
term hazard maps of the expected rate of occurrence of earthquakes are
reasonably accurate. But is it possible to do better? Are there
precursory phenomena that will allow earthquakes to be forecast?

It is actually quite surprising that immediate local precursory
phenomena are not seen. Prior to a volcanic eruption, increases in
regional seismicity and surface movements are generally observed. For
a fault system, the stress gradually increases until it reaches the
frictional strength of the fault and a rupture is initiated. It is
certainly reasonable to hypothesize that the stress increase would
cause increases in background seismicity and aseismic slip. In order
to test this hypothesis the Parkfield Earthquake Prediction Experiment
was initiated in 1985. The expected Parkfield earthquake occurred
beneath the heavily instrumented region on 28 September 2004. No local
precursory changes were observed \citep{Lindh05}.

In the absence of local precursory signals, the next question is
whether broader anomalies develop, and in particular whether there is
anomalous seismic activity. It is this question that is addressed in
this paper. Using a technique that has been successfully applied to
the forecasting of El Ni\~no we have developed a systematic pattern
informatics (PI) approach to the identification of regions of
anomalous seismic activity. Applications of this technique to
California, Japan, and on a world-wide basis have successfully
forecast the location of future earthquakes. It must be emphasized
that this is not an earthquake prediction. It is a forecast of where
future earthquakes are expected to occur during a relatively long time
window of ten years. The objective is to reduce the possible future
sites of earthquakes relative to a long term hazard assessment map.

Examination of the ROC diagrams indicates that the most important and
useful of the suite of forecast maps are those with the least number
of hotspot boxes, {\it i.e.\/}, those with small values of the false
alarm rate, $F$.  A relatively high proportion of these hotspot boxes
represent locations of future large earthquakes, however these maps
also have a larger number of failures-to-forecast.  Exactly which
forecast map(s) to be used will be a decision for policy-makers, who
will be called upon to balance the need for few false alarms against
the desire for the least number of failures-to-forecast.

Finally, we remark that the methods used to produce the forecast maps
described here can be extended and improved.  We have found
modifications to the procedures described in Section 5 that allow the
PI map to substantially outperform the RI map as indicated by the
respective ROC diagrams.  These methods are based on the approach of:
1) starting with the RI map, and introducing improvements using the
steps described for the PI method; and 2) introducing an additional
averaging step.  This new method and its results will be described in
a future publication.


\begin{acknowledgements}
This work has been supported by NASA Headquarters under the Earth
System Science Fellowship Grant NGT5 (JRH), by a JSPS Research
Fellowship (KZN), by an HSERC Discovery grant (KFT), by a grant from
the US Department of Energy, Office of Basic Energy Sciences to the
University of California, Davis DE-FG03-95ER14499 (JRH and JBR), and
through additional funding from NSF grant ATM-0327558 (DLT) and the
National Aeronautics and Space Administration under grants through the
Jet Propulsion Laboratory to the University of California, Davis.
\end{acknowledgements}



\end{document}